\documentstyle[aps,epsf]{revtex}
\draft
\preprint{MKPH-T-98-17}
\title{%
Rescattering effects in coherent $\eta$-photoproduction on the 
deuteron\thanks{Supported
by the Deutsche Forschungsgemeinschaft (SFB 201)}}
\author{Frank Ritz
  and Hartmuth Arenh\"ovel}
\address{Institut f\"ur Kernphysik, Johannes Gutenberg-Universit\"at, 
D-55099 Mainz, Germany}
\date{\today}
\begin{document}
\maketitle
\begin{abstract}
Rescattering effects in coherent eta photoproduction on the deuteron are 
studied in the region of the $S_{11}(1535)$ resonance. 
For the elementary reaction mechanism on the nucleon, an effective isobar 
model with inclusion of resonances and Born terms is used. The resonance 
parameters are fixed by considering pion photoproduction 
only. The predicted cross sections for eta production on the proton are 
in satisfactory agreement with experiment without any additional parameter 
fitting. For the coherent eta production on the deuteron we have considered 
the impulse approximation taking into account the $P_{11}(1440)$, 
$D_{13}(1520)$, and $S_{11}(1535)$ resonances, 
rescattering,
and meson exchange currents. 
For the rescattering mechanism a coupled channel model 
including the dominant $S_{11}(1535)$ resonance is used. A reduction 
of the total cross section up to about 50~\% through rescattering is found. 
This is considered as an upper limit for the size of rescattering effects 
due to uncertainties in the $NS_{11}$ 
transition potential. Other two-body effects like meson exchange currents 
appear to be small. 

\end{abstract}


\section{Introduction}
The photoproduction of $\eta$-mesons on deuterium is an extremely 
interesting process. While the incoherent reaction is studied with the main 
motivation to get information on the neutron amplitude in quasifree 
kinematics, the coherent process is considered as an isospin filter 
in order to study the isoscalar properties of $(I=1/2)$-nucleon 
resonances. Moreover, 
it offers the possibility to study rescattering and other medium effects. 
In view of quark model predictions for a small ratio $R^{(s/p)}$ of 
isoscalar over proton amplitude in conjunction with a further suppression 
due to a large energy-momentum mismatch, a very small total cross section 
for the coherent process compared to the incoherent one was expected. 
Thus it was quite a surprise that the first data of Anderson and Prepost 
\cite{AnP69} were considerably above this simple estimate 
and seemed to indicate either a larger ratio or strong rescattering effects 
as estimated by Hoshi et al.\ \cite{Hoshi}. This somewhat contradictory 
situation has changed in recent years. Firstly, more reliable though still  
approximate evaluations within the Kerman-McManus-Thaler formalism \cite{LT1} 
gave much smaller rescattering effects than the one of \cite{Hoshi}. 
Secondly, new data became available \cite{HR97} which are much lower than 
the data of \cite{AnP69} though still somewhat higher than expected from 
the small ratio $R^{(s/p)}$ as extracted from new data of incoherent 
$\eta$-production on deuterium \cite{HR97,Kru95a} which were also analyzed 
in \cite{FiA97}. 

For these reasons it appears timely to study rescattering and other 
two-body effects in a more reliable way. The present treatment differs 
from previous work in essentially two aspects. First of all, we do not use 
an effective Lagrangian model for the elementary eta production amplitude, 
fixing the undetermined parameters by fitting experimental 
$\eta$-production data. Rather we use a simple, separable resonance model 
for the coupled system of $\pi N\rightarrow\pi N$, $\pi N\rightarrow\pi\pi N$,
and 
$\pi N\rightarrow\eta N$ 
reactions as developed by Bennhold and Tanabe 
\cite{BT91} where the resonance parameters are fitted to pion-nucleon 
scattering and pion photoproduction only. Having thus fixed the model 
parameters, one can make predictions for eta photoproduction on 
the proton. This is described in Sect.~\ref{elem}. Secondly, rescattering 
effects are treated for the first time in a coupled channel model. 
Furthermore, we also include static two-body $\pi$- and $\eta$-meson 
exchange current contributions. 
This is described in Sect.~\ref{coh} where also the results are presented. 
We close with a short summary and outlook.

\section{Elementary $\pi$- and $\eta$-meson photoproduction}
\label{elem}
Instead of viewing the resonance parameters entering the elementary 
$\gamma N\rightarrow\eta N$ amplitude as free parameters to be
fitted to experimental observables of eta photoproduction, we would like to 
see, whether one can fix all parameters entering our calculation beforehand. 
To this end, we first will consider the process
$\gamma{}N\rightarrow\pi{}N$ with the intention to fix the electromagnetic
(e.m.) parameters of the resonances, which we will treat in 
a simple separable resonance model as used previously 
by Bennhold and Tanabe \cite{BT91}, henceforth referred to as BT\@. 
Within this model the coupling of the different physical channels 
$\pi{}N\rightarrow\pi{}N$, $\pi{}N\rightarrow\eta{}N$, and
$\pi{}N\rightarrow\pi\pi{}N$ is generated effectively by the self energy 
$\Sigma(W)$ of the resonances, that is through the evaluation 
of $\pi{}N$ and $\eta{}N$ loops. 
A genuine coupled channel calculation, as e.g.\ in \cite{HS98},
would be of course much more ambitious but also much more involved. 
Nonetheless, in view of the fact that one usually has to apply certain 
approximations when incorporating the elementary amplitudes into a nuclear 
system, we have refrained from a coupled channel approach in the hope that 
this simple BT model already contains the essential features of the underlying
dynamics.

We take directly the hadronic parameters from \cite{BT91} and thus their
description of the $\pi{}N\rightarrow\pi{}N$, 
$\pi{}N\rightarrow\eta{}N$, and $\pi{}N\rightarrow\pi\pi{}N$ reactions. 
But we will not take the e.m.\ vertices parametrized in \cite{BT91}, simply
because one cannot separate the resonant contribution from the non-resonant
background and we would like to have the possibility to adjust the e.m.\
pionic born terms in order to fit the experimental pion production multipoles.
Therefore, we take the standard Lagrangians, as e.g.\ in \cite{Muko1},
and derive the non-relativistic vertex functions 
\begin{eqnarray}
 v_{\gamma{}NP_{11}} &=&
 e\, \frac{g_{\gamma{}NP_{11}}}{2M_{P_{11}}}\, \mbox{i} \vec{\sigma}\times\vec{q},\\
 v_{\gamma{}ND_{13}} &=& e\, \frac{g_{\gamma{}ND_{13}}}{2M_N}\, q_0 
    \vec{\sigma}_{D_{13}N},\\
 v_{\gamma{}NS_{11}} &=& -e\, \frac{g_{\gamma{}NS_{11}}}{M_N+M_{S_{11}}}\, q_0 
    \vec{\sigma},
\end{eqnarray}
where $\vec{q}$ is the photon momentum and $q_0=|\vec{q}\,|$ the photon energy,
and repeat the fitting procedure of \cite{BT91} obtaining thus effective,
energy-dependent, and complex e.m.\ couplings. These may be viewed
as $\gamma{}N\rightarrow{}N^\ast$ form factors, reflecting the internal e.m.\
structure of the resonances. They are defined by
\begin{equation}
\label{effcoupl}
g^{(I)}_{\gamma{}NR} = \frac{X^{(I)}_{\mbox{Exp}} 
- X^{(I)}_{\mbox{Born}}}{X^{(I)}_R(g^{(I)}_{\gamma{}NR}=1)}\,,
\end{equation}
where $X^{(I)}$ denotes the $\gamma{}N\rightarrow\pi{}N$ multipole to which 
a given resonance $R$ contributes, $I=0,1$ the isospin quantum number, 
with $I=0$ corresponding to isoscalar excitation in the 
$X^{(0)}$ multipoles, and $I=1$ to the isovector 
excitation in the $X^{(1/2)}$ multipoles. $X^{(I)}_{\mbox{Exp}}$ denotes the
experimental value of the multipole taken from \cite{SAID},  
$X^{(I)}_{\mbox{Born}}$ the Born multipole, and
$X^{(I)}_{R}$ the theoretical parametrization of the resonant multipole.

\begin{figure}
\centerline{%
\epsfxsize=75.0ex
\epsffile{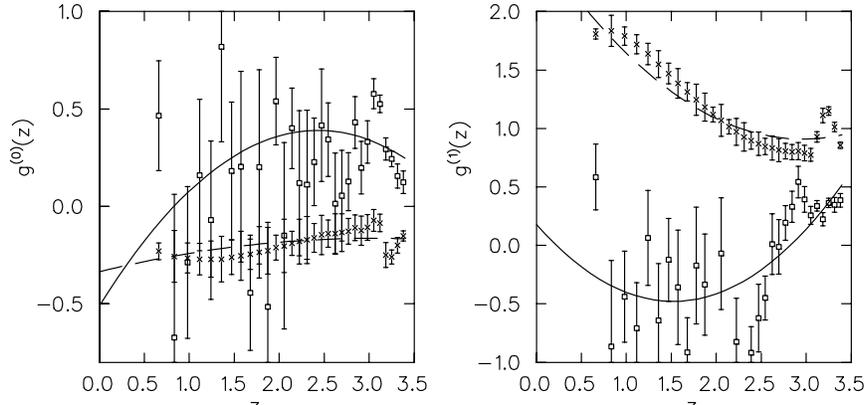}}
\caption{Effective isoscalar and isovector $\gamma{}NS_{11}$ couplings.
Notation: ($\Box$) real part, ($\times$) imaginary part, 
full curve: fit to the real part, dashed curve: fit to the imaginary part.}
\label{s11coupl}
\end{figure}

For the Born terms we use in pv-coupling  
the direct and crossed nucleon pole graphs, the
Z-graphs, Kroll-Ruderman term, pion pole terms, vertex currents (restoring
gauge invariance), and the vector meson exchange graphs. We take the 
parameters for the latter ones from \cite{EB1,GK1}.
At each $\pi{}N$ vertex we insert a form factor 
\begin{eqnarray}
 f_{\pi{}N}(k_\pi)       &=& [1+(\frac{k_\pi}{\Lambda_{\pi{}N}})^2]^{-1},
        \quad \Lambda_{\pi{}N}=600\;\mbox{MeV}\,.\label{fpi}
\end{eqnarray}
We evaluate (\ref{effcoupl}) for every discrete energy of the experimental
data set yielding thus the effective coupling first at every data point. 
Then, because we need the couplings as a function of the invariant 
mass $W$, we have fitted polynomials of the order 4 in the pion momentum
to $g^{(I)}_{\gamma{}NR}$:
\begin{equation}
g^{(I)}_{\gamma NR}(z)=\sum_{k=0}^{k=4} \alpha^{(I)}_k z^k,\quad 
z=\frac{k_\pi(W)}{m_{\pi}}\,.
\end{equation}
In Table~\ref{tablecoupl} all expansion coefficients for the $S_{11}$,
$P_{11}$, and $D_{13}$ resonances are summarized. The resulting 
effective couplings for the $\gamma{}N\rightarrow{}S_{11}$ transition 
are depicted in Fig.~\ref{s11coupl}, as well as the fitted functions. 
As one readily notices, the energy-dependence is not too strong, the 
coefficients of $z^3$ and $z^4$ are indeed very small. But it is obvious,
looking at the large error bars, 
that these fits contain a sizeable uncertainty.

\begin{figure}
\centerline{%
\epsfxsize=75.0ex
\epsffile{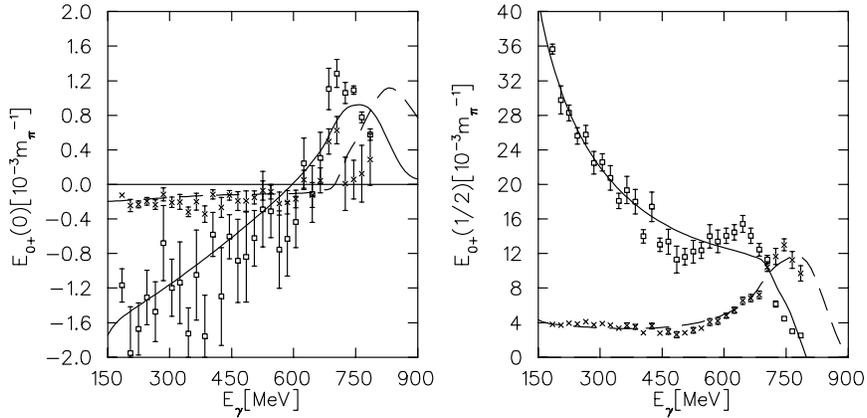}}
\caption{Fits of the $E_{0+}(0)$ and $E_{0+}(1/2)$ multipoles of pion 
photoproduction and experimental data from \protect{\cite{SAID}}. Notation: 
($\Box$) and full curve: real part, ($\times$) and dashed curve: 
imaginary part.}
\label{e0p}
\end{figure}

Now we are ready to apply the BT model to $\eta$-photoproduction. 
To this end we have to make a choice for the parameters of the Born terms. 
We have chosen a small value $g_{\eta{}N}^2/4\pi= 0.4$ (ps-coupling)
as also favoured in \cite{EB1,LT0,KT1} with the same cutoff factor as 
in (\ref{fpi}). For the vector meson couplings we have taken 
the same values as in \cite{EB1,GK1}. First we show in Fig.~\ref{e0p}
the $E_{0+}$ multipole of pion photoproduction. 
The description is quite 
satisfactory, but not as good as in \cite{BT91}, which might be due to 
the fact that our ansatz for the e.m.\ vertices and couplings is 
not as flexible as the one in \cite{BT91}.

\begin{figure}
\centerline{%
\epsfxsize=80.0ex
\epsffile{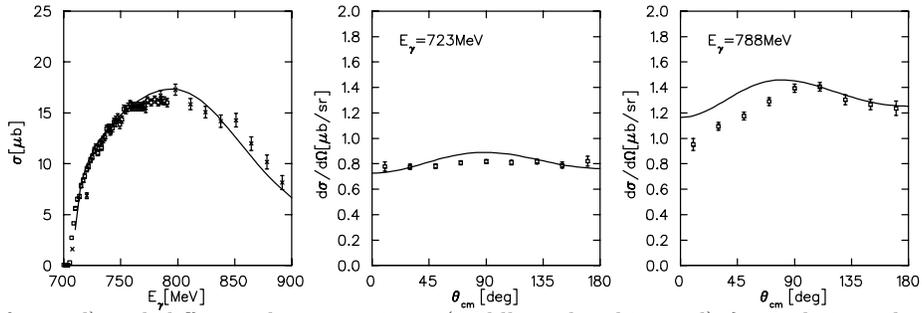}}
\caption{Total (left panel) and differential cross sections (middle and 
right panel) for $\eta$-photoproduction on the proton.
Experimental data: ($\Box$) from 
\protect{\cite{Kru95}}, ($\times$) from \protect{\cite{Wil93}}.}
\label{sigmadiffeta}
\end{figure}

The predictions for the elementary $\eta$-production 
process are shown in Fig.~\ref{sigmadiffeta} for the total and 
differential cross sections. One readily notices that the experimental 
data are described quite well, although not perfectly. We overestimate 
the maximum of the total cross section by $\sim$~5~\% which we do not 
consider as serious for the present investigation. Thus the satisfactory 
overall description is quite a success in view of the fact, that no 
additional fitting to $\eta$-photoproduction data has been applied.
A similar successful description using the BT model
including an e.m.\ background has been reached by Tiator et al.\ \cite{LT0} 
before.
All in all, we have obtained a realistic description of the elementary 
process that is simple enough to be incorporated into the deuteron. 
However, we would like to mention, that the ratio 
$|g^s_{\gamma{}NS_{11}}/g^p_{\gamma{}NS_{11}}|$ extracted from the 
present model is $\sim$~0.2, and even somewhat larger in the 
threshold region.
The comparison of these couplings with the results of other 
approaches, like the effective Lagrangian approach, 
is somehow difficult, because ours are complex, with sizeable
imaginary parts, and energy-dependent.
Nonetheless, this ratio is about the same as was needed to 
fit the data on coherent $\eta$-production on the deuteron in the 
impulse approximation by Hoffmann-Rothe et al.\  \cite{HR97}. 
A smaller value of $\sim$~0.09 was extracted from the incoherent reaction 
by the same authors assuming a simple quasifree picture while in 
\cite{FiA97} a value of $\sim$~0.11 was found.

Another remark is in order, namely the results for $\eta$-photoproduction  
presented above are still model dependent, because we found that the 
observables depend sizeably on the cutoff parameter of the pionic Born 
background used in the
BT fitting procedure. This is expected, because the channels 
$\gamma{}N\rightarrow\pi{}N$ and $\gamma{}N\rightarrow\eta{}N$ are of course 
not independent. They both would emerge from a consistent treatment of the
e.m.\ interaction within a coupled channel model, 
although this connection is somehow indirect within the BT approach.
However, the fact, that it is possible to choose the Born 
background for the pionic reaction such that the $\eta$-process is 
described simultaneously, is in our opinion
at least a consistency check of the BT model.

\section{Coherent photoproduction of $\eta$-mesons on the deuteron}
\label{coh}

\begin{figure}
\centerline{%
\epsfxsize=80.0ex
\epsffile{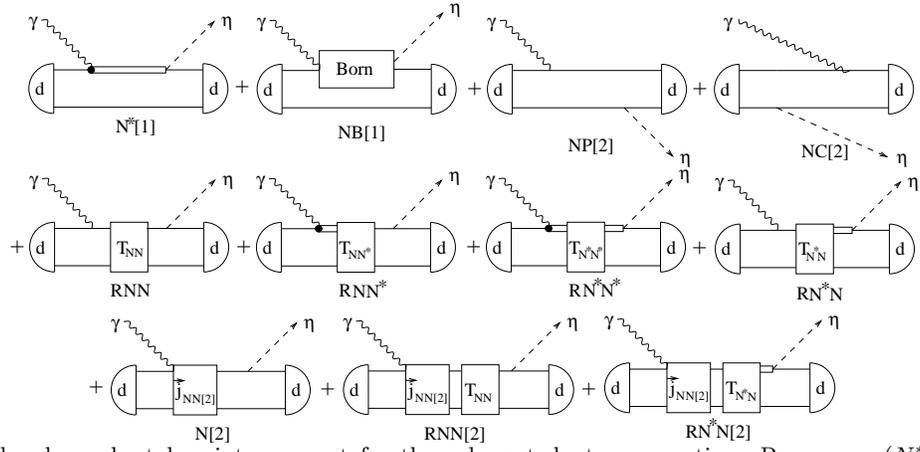}}
\caption{Time-ordered graphs taken into account for the coherent deuteron
reaction: Resonance ($N^\ast[1]$), Born ($NB[1]$), disconnected two-body terms 
($NP[2]$ and $NC[2]$), rescattering terms ($RNN$, $RNN^\ast$, 
$RN^\ast{}N^\ast$, $RN^\ast{}N$) and MEC contributions ($N[2]$, $RNN[2]$ and $RN^\ast{}N[2]$).}
\label{F1}
\end{figure}

We show in Fig.~\ref{F1} the graphs of the coherent process which we have 
included in the present calculation. They comprise the one-body resonance 
($N^\ast[1]$), Born ($NB[1]$), disconnected two-body terms ($NP[2]$ and 
$NC[2]$), various rescattering terms ($RNN$, $RNN^\ast$, $RN^\ast{}N^\ast$, $RN^\ast{}N$) 
and meson exchange current (MEC)
 contributions ($N[2]$, $RNN[2]$ and $RN^\ast{}N[2]$). As resonances we have 
considered the three resonances most likely to affect 
$\eta$-photoproduction, i.e.\ $P_{11}(1440)$, $D_{13}(1520)$, and
$S_{11}(1535)$. The various pieces of the elementary amplitude 
can be incorporated easily into the deuteron. For the energy dependence 
of the resonance couplings and the self energies, one has to specify 
the invariant mass $W_{\mbox{sub}}$ 
of the $\gamma N$-subsystem as is discussed 
in detail in \cite{EB1}. We have taken the spectator-on-shell 
approximation, because the reaction is dominated by small Fermi momenta.

It is well known that the first two resonances $P_{11}(1440)$ and 
$D_{13}(1520)$ have almost no effect on the differential cross section of 
the elementary reaction \cite{GK1}, although polarization observables may 
show a higher sensitivity to those resonances. For this reason we have 
considered $NN\rightarrow{}NS_{11}$ rescattering only besides $NN$ 
rescattering.
In order to treat the exchange of intermediate $\pi$- and $\eta$-mesons we 
have solved a set of coupled equations for the $NN$ and $NS_{11}$ channels 
using the static Bonn OBEPQ(ABC) meson exchange potentials \cite{Mach1}
and static nucleon-resonance transition potentials including 
a box-renormalization \cite{Sauer} of the $NN$ interaction,
\begin{eqnarray}
 V_{NS_{11}\leftarrow{}NN}      &=& \frac{g_{BNS_{11}} g_{BNN}}{(2\pi)^3
   2\omega_k} \frac{\vec{\sigma}_2\cdot\vec{k}}{2M_N} G_0^{\mbox{static}} 
   + (1 \leftrightarrow 2),\\
 V_{NS_{11}\leftarrow{}NS_{11}} &=& \frac{g^2_{BNS_{11}}}{(2\pi)^3
   2\omega_k} G_0^{\mbox{static}}
   + (1 \leftrightarrow 2), 
\end{eqnarray}
with
\begin{eqnarray}
  G_0^{\mbox{static}} &=& -\frac{1}{\omega_k}+\frac{1}{\Delta{}m-\omega_k}, 
  \ \mbox{for}\ B\in\{\pi,\eta\},
\end{eqnarray}
where $\vec{k}$ is the meson momentum transferred onto the first nucleon (or
resonance), $\omega_k$ is the corresponding on-shell meson energy, 
 $\Delta{}m=M-M_{S_{11}}$ in the case of the $NS_{11}\leftarrow{}NN$
potential, and $\Delta{}m=2(M-M_{S_{11}})$ for the 
$NS_{11}\leftarrow{}NS_{11}$ potential.

In order to fix the transition strength we have studied the isoscalar 
${}^1P_1$-phase of $NN$-scattering which is expected to be most 
sensitive to the incorporation of $NS_{11}$ transition potentials, 
because the $NS_{11}$ state couples to this partial $NN$ wave 
as a s-wave. The static Bonn OBEPQ(ABC) potentials describe the real part 
of this phase quite well without this resonance and we find
no large effect of the additional $NS_{11}$ interaction, also in the isovector
scattering phases. On the other hand, the inelasticities 
cannot be described satisfactorily by our model. This is probably a 
shortcoming of the present approach. A more consistent meson-theoretical 
description of $NN$ scattering would start with a model incorporating all 
relevant resonances right from the beginning so that all resonance parameters 
could be fixed by a fit to $NN$-scattering data. 
Such a calculation is beyond the scope of the present study. 
Since we were unable to fix the strength of the $NN\rightarrow{}NS_{11}$ 
and $NS_{11}\rightarrow{}NS_{11}$ transition potentials by considering 
$NN$ scattering, we have chosen the same couplings as for the elementary 
process given in \cite{BT91}, but
use a pole-normalized cutoff of 
$\Lambda_{\pi{}NS_{11}}=\Lambda_{\eta{}NS_{11}}=1\;\mbox{GeV}$.
With this choice we believe to maximize rescattering effects.

Turning now to the results of the coherent process on the deuteron we show in 
Fig.~\ref{sigmadiffetadeut} the theoretical total cross section for coherent
$\eta$-production on the deuteron. Evidently, the reaction is dominated by
the resonance contribution. The Born terms give a small reduction at lower
energies but lead to an enhancement near the maximum shifting it slightly to
higher energies. Rescattering effects yield a sizeable reduction, in
particular very close to the threshold where they dominate.
The additional MEC contributions are small, they amount to $\sim$~3~\%.
Furthermore, the 
first order rescattering, i.e., the one-loop approximation $T=V$ is a 
reasonable approximation for the coherent reaction only close to 
threshold. 
It breaks down very soon for higher energies, 
leading to an enhancement above
$E_\gamma=750\;\mbox{MeV}$ in contrast to the reduction of 
a complete coupled channel calculation. 

\begin{figure}
\centerline{%
\epsfxsize=50.0ex
\epsffile{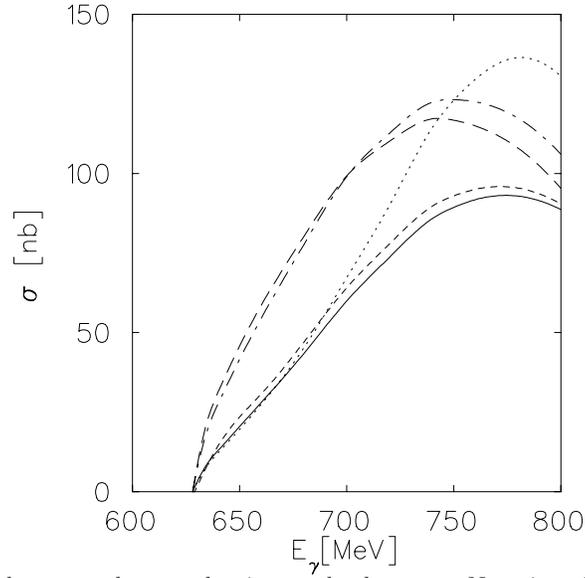}}
\caption{Total cross sections for coherent $\eta$-photoproduction 
on the deuteron.
Notation of the curves: long dashed: 
resonances in impulse approximation,
dash-dotted: Born contributions added, 
short dashed: rescattering contribution added,
dotted: with rescattering contributions in the one-loop approximation only,
full: complete calculation including rescattering and MEC\@.}
\label{sigmatotetadeut}
\end{figure}

\begin{figure}
\centerline{%
\epsfxsize=65.0ex
\epsffile{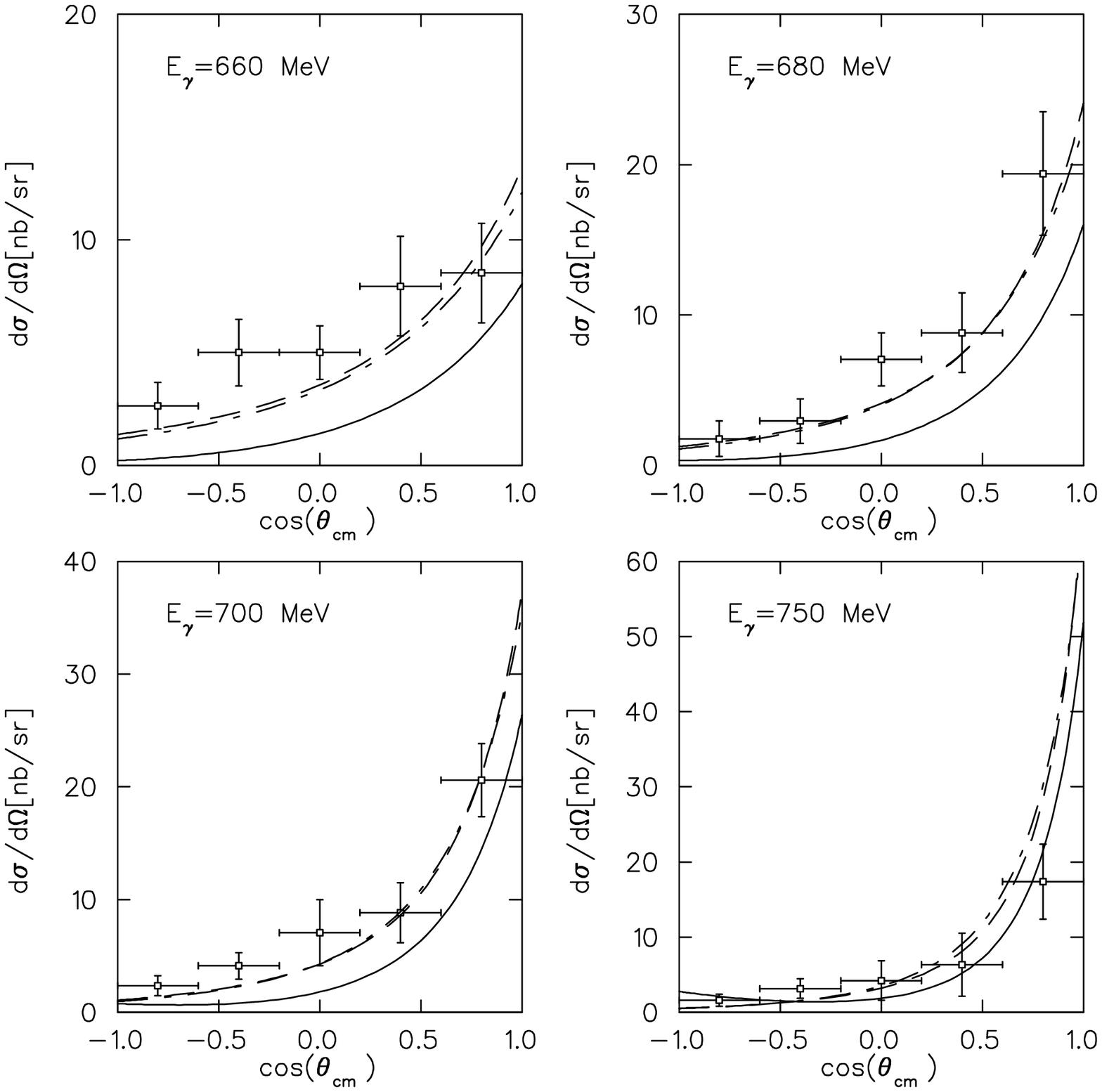}}
\caption{Differential cross sections for coherent $\eta$-photoproduction
on the deuteron.
Notation of the curves as 
in Fig.~\protect{\ref{sigmatotetadeut}},
experimental data from \protect{\cite{HR97}}.}
\label{sigmadiffetadeut}
\end{figure}

As next we show in Fig.~\ref{sigmatotetadeut}
 the data of \cite{HR97} and the theoretical 
differential cross sections in the impulse approximation and with inclusion 
of rescattering and MEC.
The impulse approximation appears to be in 
quite good agreement with the data. It is dominated by the $S_{11}$ 
resonance, the Born terms and vector meson graphs are small 
in accordance with the findings of \cite{LT1}, although single Born 
terms, like the Z-graphs or the $\omega$ meson exchange graph, 
are not negligible.
Inclusion of 
rescattering leads to a sizeable decrease, in particular in the backward
direction, while MEC contributions, not separately shown, are 
small, similar to what was found for the analogous reaction of 
$\pi^0$ production on the deuteron \cite{PW1}. Also Kamalov et al.\ found 
a decrease by rescattering, but much smaller in size, which likely is due 
to their neglect of pion exchange. 
Our findings are at variance with those of N.\ Hoshi et al.\ 
\cite{Hoshi} which were much larger in size and which lead to an increase
of the cross section. 
Especially,
we find that the effect of $\eta$-exchange is strongly 
suppressed compared to pion exchange. 
In order to have the effect of $\eta$-rescattering to be of the same 
magnitude as $\pi$-rescattering, the $\eta{}NS_{11}$ coupling would have 
to be much bigger than the $\pi{}NS_{11}$ coupling strength. But the 
partial decay widths of $S_{11}\rightarrow\pi{}N$ and 
$S_{11}\rightarrow\eta{}N$ certainly rule out this scenario.

Compared to the experimental data, one notices a systematic underestimation 
of the complete calculation 
by about 30 to 40~\% for energies below 750 MeV while at 750 MeV the 
agreement is quite satisfactory. On the other hand, we had already remarked
above 
that we consider the size of rescattering contributions as an upper 
limit, the reason for this being that we were unable to determine precisely the
magnitude of the appropriate transition potentials, 
as was the case in \cite{PW1} for the 
corresponding reaction $\gamma{}d\rightarrow\pi^0d$. Therefore, in view of 
the fact that we found with our maximal choice of parameters a reduction 
of up to 50~\% in the total cross section by rescattering, it is very likely 
that a more realistic choice of parameters would reduce the size of 
rescattering effects. In other words, the decrease of the cross 
section by rescattering might appear too strong. However, considering the 
quite large error bars it might be premature to draw a definite conclusion 
here as to whether a serious discrepancy between theory and experiment 
exists or not.

As a summary we would like to state that the most important contribution 
is the one-body part involving the $S_{11}$ resonance. Thus the 
properties of this resonance, especially the parametrization of its self 
energy, govern the coherent reaction on the deuteron. The e.m.\ structure 
of the $S_{11}$ resonance used in this work gives a realistic description 
of the elementary process and thus should be reliable also for the coherent 
production on the deuteron. 
The sum of nucleonic Born terms and vector meson contributions is small.
 Among the medium effects, the 
rescattering is sizeable and leads to a reduction of the differential cross 
section, while MEC contributions are much smaller. The role of the latter 
might change in electroproduction with increasing momentum transfers. 
With respect to the question whether a discrepancy between coherent and 
incoherent production exists, as claimed in \cite{HR97},
a convincing test would be the incorporation 
of our effective couplings into a calculation of the incoherent process 
$\gamma{}d\rightarrow\eta{}NN$. This will be the topic of a forthcoming paper.

\begin{table}
\caption{%
Parameters for the effective e.m.\ couplings of the $P_{11}$, $D_{13}$, and
$S_{11}$ resonances.%
\label{tablecoupl}}
{\scriptsize
\begin{tabular}{c|cc|cc|cc}
     & \multicolumn{2}{c|}{$S_{11}$}
 & \multicolumn{2}{c|}{$P_{11}$} & \multicolumn{2}{c}{$D_{13}$} \\
\hline
  & Re & Im & Re & Im & Re & Im \\
\hline
$\alpha^{(0)}_4$ & $2.2\cdot 10^{-6}$ & $2.5\cdot 10^{-7}$ & $1.2\cdot 10^{-7}$ 
& $-5.6\cdot 10^{-8}$ & $-4.7\cdot 10^{-6}$ & $2.5\cdot 10^{-7}$ \\

$\alpha^{(0)}_3$ & $8.7\cdot 10^{-5}$ & $8.7\cdot 10^{-6}$ & $1.1\cdot 10^{-5}$ 
& $-5.6\cdot 10^{-6}$ & $ -1.8\cdot 10^{-4}$ & $1.2\cdot 10^{-5}$ \\
 
$\alpha^{(0)}_2$ & $-0.153$ & $-1.85\cdot 10^{-2}$ & $2\cdot 10^{-4}$ 
& $-2.4\cdot 10^{-4}$ & $0.334$ & $ -1.75\cdot 10^{-2}$  \\

$\alpha^{(0)}_1$ & $0.741$ & $0.112$ & $-0.868$ & $0.436$ & $-2.322$ 
& $ 0.131$  \\

$\alpha^{(0)}_0$ & $-0.509$ & $-0.335$ & $3.043$ & $-1.584$ & $ 3.960$  
& $ -0.151$ \\
\hline
$\alpha^{(1)}_4$ & $-4.1\cdot 10^{-6}$ & $-2.7\cdot 10^{-6}$ & $ 6\cdot 10^{-8}$ 
& $2.7\cdot 10^{-6}$ & $-4.4\cdot 10^{-6}$ & $-5.1\cdot 10^{-7}$ \\

$\alpha^{(1)}_3$ & $-2.4\cdot 10^{-4}$ & $-1.3\cdot 10^{-4}$ 
& $ 5.9\cdot 10^{-6}$ & $8.4\cdot 10^{-5}$ & $-1.4\cdot 10^{-4}$ 
& $-1.8\cdot 10^{-5}$ \\

$\alpha^{(1)}_2$ & $0.284$ & $0.193$ & $1.5\cdot 10^{-4}$& $-0.185$ 
& $ 0.311$ & $3.48\cdot 10^{-2}$ \\

$\alpha^{(1)}_1$ & $-0.864$ & $-1.136$ & $-0.431$ & $ 0.924$ & $-1.850$ 
& $3.68\cdot 10^{-2}$ \\

$\alpha^{(1)}_0$ & $ 0.178$ & $2.588$ & $2.905$ & $0.295$ & $4.202$ 
& $2.35\cdot 10^{-2}$\\
\end{tabular}
}
\end{table}

\begin{references}
\bibitem{AnP69} R.L.\ Anderson and R.\ Prepost, Phys.\ Rev.\ Lett.\ 
{\bf 23} (1969) 46.

\bibitem{Hoshi} N.\ Hoshi, H.\ Hyuga, and K.\ Kubodera, 
Nucl.\ Phys.\ A {\bf 324} (1979) 234.

\bibitem{LT1} S.S.\ Kamalov, L.\ Tiator, and C.\ Bennhold,
Phys.\ Rev.\ C {\bf 55} (1997) 98.

\bibitem{HR97} P.\ Hoffmann-Rothe {\it et al.},
Phys.\ Rev.\ Lett.\ {\bf 78} (1997) 4697.

\bibitem{Kru95a} B.\ Krusche {\it et al.}, Phys.\ Lett.\ B {\bf 358} (1995) 40.

\bibitem{FiA97} A.\ Fix and H.\ Arenh\"ovel, Z.\ Phys.\ A {\bf 359} 
(1997) 427.


\bibitem{BT91} C.\ Bennhold and H.\ Tanabe, Nucl.\ Phys.\ A {\bf 530}
 (1991) 625.

\bibitem{HS98} C.\ Sch\"utz, J.\ Haidenbauer, J.\ Speth, and
J.W.\ Durso, Phys.\ Rev.\ C {\bf 57} (1998) 1464.

\bibitem{Muko1} N.C.\ Mukhopadhyay, J.F.\ Zhang, and M.\ Benmerrouche,
Phys.\ Lett.\ B {\bf 364} (1995) 1.

\bibitem{SAID} R.\ Arndt {\it et al.}, 
 program ``SAID'', URL http://said.phys.vt.edu,
 multipole solution ``SP97''.


\bibitem{EB1} E.\ Breitmoser and H.\ Arenh\"ovel,
Nucl.\ Phys.\ A {\bf 612} (1997) 321.

\bibitem{GK1} G.\ Kn\"ochlein, D.\ Drechsel, L.\ Tiator,
 Z.\ Phys.\ A {\bf 352} (1995) 327.

\bibitem{LT0} L.\ Tiator, C.\ Bennhold, and S.S.\ Kamalov,
Nucl.\ Phys.\ A {\bf 580} (1994) 455.

\bibitem{KT1} M.\ Kirchbach and L.\ Tiator,
Nucl.\ Phys.\ A {\bf 604} (1996) 385.


\bibitem{Mach1} R.\ Machleidt, K.\ Holinde, and Ch.\ Elster,
Phys.\ Rep.\ {\bf 149} (1987) 1; 
R.\ Machleidt, Adv.\ Nucl.\ Phys.\ {\bf 19} (1989) 189.

\bibitem{Sauer} H.\ P\"opping, P.U.\ Sauer, and X.-Z.\ Zhang, 
Nucl.\ Phys.\ A {\bf 474} (1987) 557.

\bibitem{Wil93} M.\ Wilhelm, PhD thesis, Universit\"at Bonn, 1993.

\bibitem{Kru95} B.\ Krusche {\it et al.},
Phys.\ Rev.\ Lett.\ {\bf 74} (1995) 3736.

\bibitem{PW1} P.\ Wilhelm and H.\ Arenh\"ovel,
Nucl.\ Phys.\ A {\bf 609} (1996) 469.

\end{references}
\end{document}